\documentclass[aps,prd,reprint]{revtex4-2}
\usepackage{hyperref}
\usepackage{graphicx}
\usepackage{commath}
\usepackage{bm}
\usepackage{amsfonts}
\begin{document}
\title{Topological Stabilization via Higgs and \textit{Z}-Boson Mediated Repulsions in Electroweak Monopole-Antimonopole Pairs}
\author{Dan Zhu and Xurong Chen}
\affiliation{Southern Center for Nuclear Science Theory, Institute of Modern Physics, Chinese Academy of Sciences, Huizhou, 516000, China\\School of Nuclear Science and Technology, University of Chinese Academy of Sciences, Beijing, 100049, China}
\author{Khai-Ming Wong}
\affiliation{School of Physics, Universiti Sains Malaysia, 11800 USM, Penang, Malaysia}

\begin{abstract}
We identify two distinct repulsive mechanisms in the Cho-Maison monopole-antimonopole pair (MAP) configuration. Our results show that the Higgs-mediated repulsion exhibits a non-monotonic dependence on both topological charge and Higgs self-coupling, confirming its topological origin while revealing a mass-controlled range transition that deviates from the exponential form of a Yukawa potential. Simultaneously, the $Z$-boson field generates localized repulsive cores of radius $R_c\approx0.8\,m_W^{-1}$, consistent with the weak interaction scale. The collaborative effect of these mechanisms---operating in different physics regimes---counteracts the magnetic attraction, establishing a stabilization paradigm for the Cho-Maison MAP that extends naturally to other topological solitons in the Standard Model and various systems described by effective field theories.
\end{abstract}
\maketitle

\section{Introduction}
Since the seminal discovery of the 't Hooft-Polyakov monopole \cite{tHooft,Polyakov}, these topological defects have become essential to the grand unified theories, where their existence is fundamentally required \cite{Preskill}. Beyond particle physics, they provide critical insights into non-perturbative phenomena including cosmological phase transitions \cite{Preskill}, color confinement mechanisms \cite{SW}, monopole-catalyzed baryogenesis \cite{Rubakov,Callan}, etc.

Following Taubes' rigorous proof of existence for static monopole-antimonopole pairs (MAPs) in the SU(2) Yang-Mills-Higgs theory \cite{Taubes}, their first numerical realization was achieved by Kleihaus and Kunz \cite{KKMAP}. Subsequent studies on the scattering, pair creation \cite{VachaspatiPRL}, and interaction potentials \cite{MP1,MP2} revealed rich dynamics of such systems. The MAP configuration serves as an ideal platform for probing force equilibria where long-range topological attraction competes against short-range repulsive interactions. However, these studies remained fundamentally limited by their SU(2) symmetry—incapable of describing electroweak-scale physics where the U(1)$_Y$ gauge field critically modifies topological interactions.

Contrary to the expectation that the second homotopy $\pi_2[\text{SU(2)}_L\times\text{U(1)}_Y/\text{U(1)}_\text{em}]=0$ precludes magnetic monopoles in the Weinberg-Salam model, Cho and Maison demonstrated its \textit{CP}$^1$ interpretation \cite{ChoMaison}. The Weinberg-Salam theory is viewed as a gauged \textit{CP}$^1$ model in which the normalized Higgs doublet plays the role of the \textit{CP}$^1$ field, endowing the SU(2) sector with $\pi_2\left(S^2\right)=\mathbb{Z}$ topology identical to the Georgi-Glashow model \cite{ChoPhilTrans}. This breakthrough legitimized electroweak monopoles theoretically and enabled the construction of Cho-Maison MAP solutions \cite{ZhuDan2}.

In this work, we study the monopole-antimonopole interactions within the Cho-Maison MAP configuration by means of a detailed numerical analysis of the stress-energy tensor $T_{\mu\nu}$, validated against the well-established SU(2) MAP solutions \cite{KKMAP}. Our results indicate that repulsive interactions mediated by the Higgs and $Z$-boson fields compete with the topological magnetic attraction, thereby resisting annihilation.

Specifically, the Higgs-mediated repulsion possesses a double non-monotonic dependence on topological charge and Higgs boson mass, revealing a complexity beyond the exponential decay of a Yukawa potential. Simultaneously, the $Z$-boson repulsion generates localized cores of radius $R_c\approx0.8\,m_W^{-1}$, directly reflecting the weak interaction scale. Together, these mechanisms establish a paradigm for topological stabilization that is applicable to other non-perturbative configurations in the Standard Model and to diverse systems described by effective field theories.

\section{Theoretical Frameworks}
We work in flat spacetime with the mostly-positive metric, ($-$+++). The following subsections detail the two underlying field theories.

\subsection{The Weinberg-Salam Model}
The Cho-Maison MAP numerical solutions are obtained in the bosonic sector of the Weinberg-Salam model, described by the following Lagrangian \cite{ChoMaison,ZhuDan2}
	\begin{align}
	\mathcal{L}=&-\frac{1}{4}F^a_{\mu\nu}F^{a\mu\nu}-\frac{1}{4}f_{\mu\nu}f^{\mu\nu}\nonumber\\
	&-\left(\mathcal{D}_\mu\bm\phi\right)^\dagger\left(\mathcal{D}^\mu\bm\phi\right)-\frac{\lambda}{2}\left(\bm\phi^\dagger\bm\phi-\frac{\mu_H^2}{\lambda}\right)^2,
	\label{eqn:LagrangianWS}
	\end{align}
with the SU(2)$\times$U(1) covariant derivative
	\begin{equation}
	\mathcal{D}_\mu=\partial_\mu-\frac{i}{2}gA^a_\mu\sigma^a-\frac{i}{2}g'B_\mu.\label{eqn:curlyDmu}
	\end{equation}
Here, $F^a_{\mu\nu}$, $A^a_\mu$, and $g$ are the SU(2) field strength, gauge potential, and coupling constant, respectively, while their U(1) counterparts are denoted $f_{\mu\nu}$, $B_\mu$, and $g'$. The spacetime indices $\mu,\nu$ range from 0 to 3.

Variations of the action for Eq. \ref{eqn:LagrangianWS} yields the stress-energy tensor and the equations of motion:
\begin{align}
	T_{\mu\nu}&=2\left(\mathcal{D}_\mu\bm\phi\right)^\dagger\left(\mathcal{D}_\nu\bm\phi\right)+F^{a\beta}_\mu F^a_{\nu\beta}+f^\beta_\mu f_{\nu\beta}\nonumber\\
	&\quad+g_{\mu\nu}\mathcal{L},\label{eqn:TmunuWS}\\
	D^\mu F^a_{\mu\nu}&=\frac{ig}{2}\left[\bm\phi^\dagger\sigma^a\left(\mathcal{D}_\nu\bm\phi\right)-\left(\mathcal{D}_\nu\bm\phi\right)^\dagger\sigma^a\bm\phi\right],\\
	\partial^\mu f_{\mu\nu}&=\frac{ig'}{2}\left[\bm\phi^\dagger\left(\mathcal{D}_\nu\bm\phi\right)-\left(\mathcal{D}_\nu\bm\phi\right)^\dagger\bm\phi\right],\\
	\mathcal{D}^\mu\mathcal{D}_\mu\bm\phi&=\lambda\left(\bm\phi^\dagger\bm\phi-\frac{\mu_H^2}{\lambda}\right)\bm\phi.\label{eqn:EoMs}
\end{align}

\subsection{The Yang-Mills-Higgs Theory}
For comparison, the SU(2) MAP configuration is constructed in the Yang-Mills-Higgs theory, governed by the Lagrangian shown below \cite{Amin}
\begin{align}
	\mathcal{L}=&-\frac{1}{4}F^a_{\mu\nu}F^{a\mu\nu}-\frac{1}{2}D_\mu\Phi^aD^\mu\Phi^a\nonumber\\
	&-\frac{\lambda}{4}\left(\Phi^a\Phi^a-\frac{\mu_H^2}{\lambda}\right)^2,\label{eqn:LagrangianYMH}
\end{align}
where the SU(2) covariant derivative is
\begin{equation}
	D_\mu=\partial_\mu+g\varepsilon^{abc}A^b_\mu.\label{eqn:YMHD}
\end{equation}
The corresponding stress-energy tensor and equations of motion are
\begin{align}
	T_{\mu\nu}&=D_\mu\Phi^aD_\nu\Phi^a+F^a_{\mu\beta}F^{a\beta}_\nu+g_{\mu\nu}\mathcal{L},\label{eqn:TmunuYMH}\\
	D^\mu F^a_{\mu\nu}&=g\varepsilon^{abc}\Phi^bD_\nu\Phi^c,\\
	D^\mu D_\mu\Phi^a&=\lambda\Phi^a\left(\Phi^b\Phi^b-\frac{\mu_H^2}{\lambda}\right).
\end{align}

In this theory, the Higgs field is an isotriplet $\Phi^a$, whereas in the Weinberg-Salam model, it is a complex scalar doublet $\bm\phi$ as shown in Eq. \ref{eqn:LagrangianWS}. In both cases, $\mu_H$ and $\lambda$ represent the Higgs field mass and self-coupling, respectively.

\section{Magnetic Ansatz}
Solutions to the equations of motion corresponding to the MAP configurations are obtained when the fundamental fields adopt the following form \cite{ZhuDan2,Amin} (spatial index $i=1,2,3$):
	\begin{align}
	gA^a_i=&-\frac{1}{r}\psi_1(r,\theta)\hat{n}^a_\phi\hat{\theta}_i+\frac{n}{r}\psi_2(r,\theta)\hat{n}^a_\theta\hat{\phi}_i\nonumber\\
	&+\frac{1}{r}R_1(r,\theta)\hat{n}^a_\phi\hat{r}_i-\frac{n}{r}R_2(r,\theta)\hat{n}^a_r\hat{\phi}_i,\nonumber\\
	A_0=&\;0.\label{eqn:AnsatzSU2}\\
	g'B_i=&\ \frac{n}{r\sin\theta}B_s(r,\theta)\hat{\phi}_i,\nonumber\\
	B_0=&\;0.\\
	\Phi^a=&\ \Phi_1(r,\theta)\hat{n}^a_r+\Phi_2(r,\theta)\hat{n}^a_\theta=H(r,\theta)\hat{\Phi}^a,\label{eqn:AnsatzHiggs1}
	\end{align}
where $H(r,\theta)=\abs{\Phi}=\sqrt{\Phi_1^2+\Phi_2^2}$ is the Higgs modulus and the corresponding unit vector is
	\begin{align}
	\hat{\Phi}^a=&-\bm\xi^\dagger\sigma^a\bm\xi\nonumber\\
	=&\cos(\alpha-\theta)\hat{n}^a_r+\sin(\alpha-\theta)\hat{n}^a_\theta=h_1\hat{n}^a_r+h_2\hat{n}^a_\theta,\nonumber\\
	\bm\xi=&\ i
		\begin{pmatrix}
		\sin\frac{\alpha(r,\theta)}{2}e^{-in\phi}\\
		-\cos\frac{\alpha(r,\theta)}{2}
		\end{pmatrix}.\label{eqn:AnsatzHiggs2}
	\end{align}
The Higgs field orientation in the SU(2) internal space is parameterized by the angle $\alpha(r,\theta)$, whose asymptotic form $\alpha\rightarrow p\theta$ \cite{Teh} is set by the integer $p$. This parameter specifies the number of poles, which is set to $p=2$ (corresponding to MAPs) in this study.

The implementation of the magnetic ansatz differs between the two theories. The Cho-Maison MAP in the Weinberg-Salam model utilizes the full ansatz, whereas for its SU(2) counterpart, only Eqs. \ref{eqn:AnsatzSU2} \& \ref{eqn:AnsatzHiggs1} are required. Additionally, the spherical coordinate system unit vectors in Eq. \ref{eqn:AnsatzSU2} are defined as
	\begin{align}
	\hat{r}_i&=\sin\theta\cos\phi\,\delta_{i1}+\sin\theta\sin\phi\,\delta_{i2}+\cos\theta\,\delta_{i3},\nonumber\\
	\hat{\theta}_i&=\cos\theta\cos\phi\,\delta_{i1}+\cos\theta\sin\phi\,\delta_{i2}-\sin\theta\,\delta_{i3},\nonumber\\
	\hat{\phi}_i&=-\sin\phi\,\delta_{i1}+\cos\phi\,\delta_{i2},
	\end{align}
and the unit vectors for isospin coordinate system are given by 
	\begin{align}
	\hat{n}^a_r&=\sin\theta\cos n\phi\,\delta^a_1+\sin\theta\sin n\phi\,\delta^a_2+\cos\theta\,\delta^a_3,\nonumber\\
	\hat{n}^a_\theta&=\cos\theta\cos n\phi\,\delta^a_1+\cos\theta\sin n\phi\,\delta^a_2-\sin\theta\,\delta^a_3,\nonumber\\
	\hat{n}^a_\phi&=-\sin n\phi\,\delta^a_1+\cos n\phi\,\delta^a_2,
	\end{align}
where $n$ denotes the $\phi$-winding number that controls the topological charge carried by each pole.

\section{Numerical Method}
The magnetic ansatz is substituted into the equations of motion, reducing the Weinberg-Salam system to seven coupled nonlinear partial differential equations, while the corresponding SU(2) system yields six. To facilitate the numerical calculations, we introduce the following dimensionless variables:
	\begin{equation}
	x=m_Wr,\;\widetilde{H}=\frac{H}{H_0},\;\tan\theta_W=\frac{g'}{g},\;\beta^2=\frac{\lambda}{g^2},
	\end{equation}
where $m_W$ is the $W$-boson mass and $H_0$ the Higgs vacuum expectation value. These quantities are defined differently in the two theories: in the SU(2) theory, $m_W=gH_0$ and $H_0=\mu_H/\sqrt{\lambda}$, whereas in the Weinberg-Salam model, $m_W=gH_0/2$ and $H_0=\sqrt{2}\mu_H/\sqrt{\lambda}$. In addition, 1 $m_W^{-1}$ is the characteristic range of weak interactions, which is roughly 2.5 attometers ($10^{-18}$m).

The dimensionless radial coordinate $x$ is compactified via $x_c=x/(x+1)$, mapping the semi-infinite domain $r\in[0,\infty)$ to $x_c\in[0,1]$. This compactification scheme naturally increases node density near the origin upon discretization, thereby enhancing the resolution in regions with the most field variations. The polar coordinate $\theta\in[0,\pi]$ is kept unchanged.

After the dimensionless transformation, the Weinberg-Salam model is fully characterized by two running parameters: the weak mixing angle $\tan\theta_W$ and the rescaled Higgs self-coupling $\beta$. Using the measured values $m_H=125.10$ GeV, $m_W=80.379$ GeV, $m_Z=91.1876$ GeV \cite{PDG}, together with the relations $\beta=m_H/(2m_W),\cos\theta_W=m_W/m_Z$, we obtain $\beta=0.7782$ and $\tan\theta_W=0.5356$. In the SU(2) theory, only the parameter $\beta$ is present after the transformation.

The system of equations is then discretized onto a non-equidistant grid of $70\times60$ using a second-order finite difference method. At the boundaries, forward/backward differences are applied, while central differences are used on the interior. The resulting nonlinear algebraic equations are solved iteratively using a trust-region reflective algorithm supplied with a Jacobian sparsity pattern to accelerate convergence. 

Boundary conditions for the Cho-Maison MAP configuration are imposed as follows \cite{ZhuDan2}. Along the positive and negative $z$-axis, when $\theta=0$ and $\pi$,
	\begin{align}
	\partial_\theta\psi_A=R_A=\partial_\theta\Phi_1=\Phi_2=\partial_\theta B_s=0,\label{eqn:BCtheta}
	\end{align}
where $A=1$, 2. Asymptotically,
	\begin{align}
	\psi_A(\infty,\theta)&=2,R_A(\infty,\theta)=B_s(\infty,\theta)=0,\nonumber\\
	\Phi_1(\infty,\theta)&=\cos\theta,\Phi_2(\infty,\theta)=\sin\theta,\label{eqn:BCrinf}
	\end{align}
and at the origin, 
	\begin{align}
	&\psi_A(0,\theta)=R_A(0,\theta)=0,B_s(0,\theta)=-2,\nonumber\\
	&\Phi_1(0,\theta)\sin\theta+\Phi_2(0,\theta)\cos\theta=0,\nonumber\\
	&\partial_r(\Phi_1(r,\theta)\cos\theta-\Phi_2(r,\theta)\sin\theta)|_{r=0}=0.\label{eqn:BCr0}
	\end{align}
For the SU(2) MAP solutions, the same set of boundary conditions applies \cite{Amin} except that there is no $B_s$.

The obtained numerical solutions exhibit high accuracy, with the squared Euclidean norm of the residuals ranging from $10^{-18}$ to $10^{-23}$ and the first-order optimality conditions satisfied within $10^{-4}$ to $10^{-7}$. The truncation errors scale as $\mathcal{O}(\Delta x_c^2)\sim\mathcal{O}(1/70^2)$ radially and $\mathcal{O}(\Delta\theta^2)\sim\mathcal{O}(\pi^2/60^2)$ polarly.

\section{Properties of the MAP configurations}
The analysis of monopole-antimonopole interactions in both MAP solutions concentrates on the $T_{33}$ component of the stress-energy tensor. In these configurations, the monopole and antimonopole are positioned symmetrically along the $z$-axis, establishing it as the principal axis of force transmission. This cylindrical symmetry necessitates the examination of $T_{33}$.

As with a single Cho-Maison monopole, its MAP configuration possesses singularities at the monopole locations, preventing the direct evaluation of $T_{33}$. To circumvent this issue, we decompose it into contributions from the fundamental fields. In the Weinberg-Salam model, according to Eq. \ref{eqn:TmunuWS}, the decomposition reads:
\begin{align}
	T_{33}^{\text{Higgs}}&=2\left(\mathcal{D}_3\bm\phi\right)^\dagger\left(\mathcal{D}_3\bm\phi\right)-\left(\mathcal{D}_i\bm\phi\right)^\dagger\left(\mathcal{D}^i\bm\phi\right)\nonumber\\
	&\quad\;-\frac{\lambda}{2}\left(\bm\phi^\dagger\bm\phi-\frac{\mu_H^2}{\lambda}\right)^2,\nonumber\\
	T_{33}^{\text{SU(2)}}&=F^a_{3i}F^{ai}_3-\frac{1}{4}F^a_{ij}F^{aij},\nonumber\\
	T_{33}^{\text{U(1)}}&=f_{3i}f_3^i-\frac{1}{4}f_{ij}f^{ij},\nonumber\\
	T_{33}&=T_{33}^{\text{Higgs}}+T_{33}^{\text{SU(2)}}+T_{33}^{\text{U(1)}}.\label{eqn:t33}
\end{align}
Likewise in the SU(2) Yang-Mills-Higgs theory, starting from Eq. \ref{eqn:TmunuYMH},
\begin{align}
	T_{33}^{\text{Higgs}}&=D_3\Phi^aD_3\Phi^a-\frac{1}{2}D_i\Phi^aD^i\Phi^a\nonumber\\
	&\quad\;-\frac{\lambda}{4}\left(\Phi^a\Phi^a-\frac{\mu_H^2}{\lambda}\right)^2,\nonumber\\
	T_{33}^{\text{SU(2)}}&=F^a_{3i}F^{ai}_3-\frac{1}{4}F^a_{ij}F^{aij},\nonumber\\
	T_{33}&=T_{33}^{\text{Higgs}}+T_{33}^{\text{SU(2)}}.\label{eqn:t33YMH}
\end{align}

\begin{figure*}
	\includegraphics[width=\textwidth]{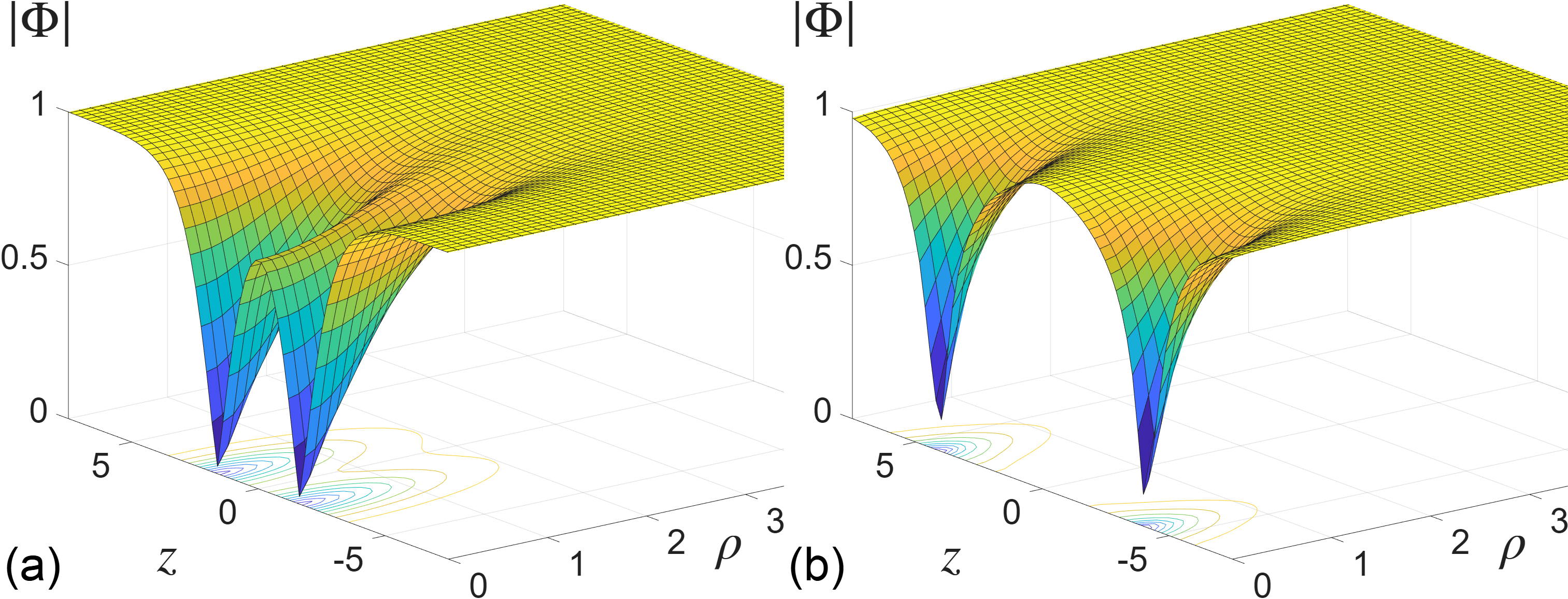}
	\caption{Higgs modulus $|\Phi|$ comparison between (a) an SU(2) MAP and (b) a Cho-Maison MAP.}
	\label{fig:HiggsModulus}
\end{figure*}

\begin{figure*}
	\includegraphics[width=\textwidth]{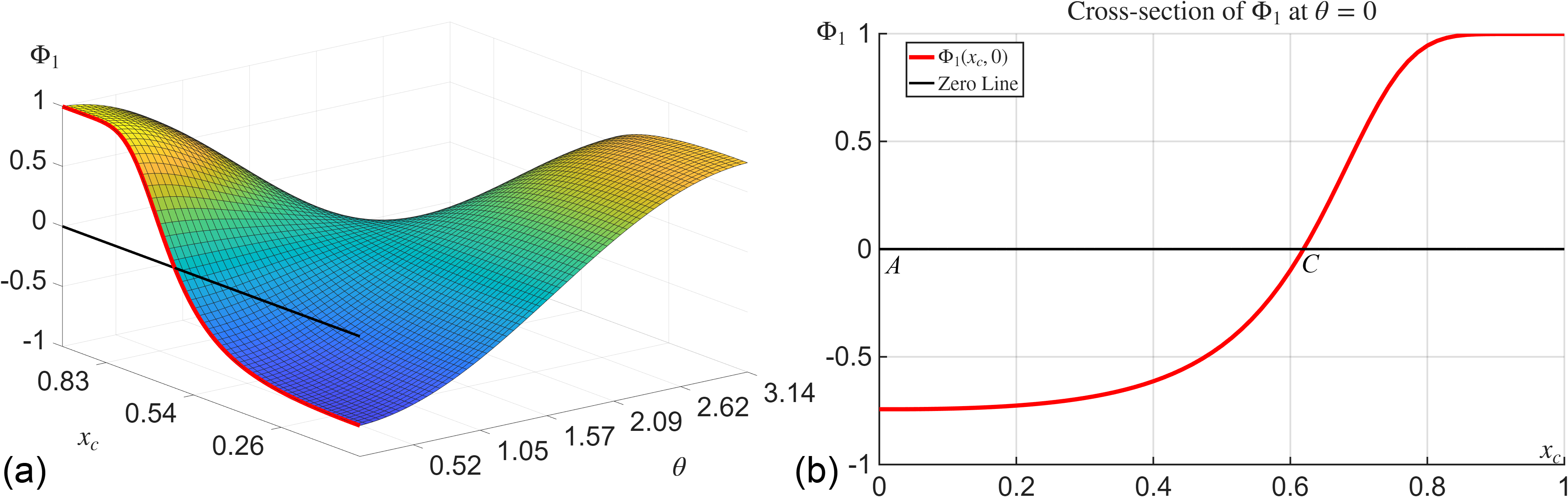}
	\caption{Determining the pole separation $d_z$ of an MAP solution from (a) $\Phi_1$ surface plot and (b) the cross-section $\Phi_1(x_c,0)$.}
	\label{fig:Phi1}
\end{figure*}

While $T_{33}$ is physical, its decomposition into individual field contributions is not. To address this, we reformulate the decomposition using observable fields. In the Weinberg-Salam model, these are defined in terms of the fundamental fields by:
	\begin{align}
		\begin{pmatrix}
		A_i^\text{em}\\
		Z_i
		\end{pmatrix}&=
		\begin{pmatrix}
		\cos\theta_W&\sin\theta_W\\
		-\sin\theta_W&\cos\theta_W
		\end{pmatrix}
		\begin{pmatrix}
		B_i\\
		A'^3_i
		\end{pmatrix},\label{eqn:WSMix}
	\end{align}
with inverse relations:
\begin{align}
B_i &= \cos\theta_WA_i^\text{em} - \sin\theta_WZ_i,\label{eqn:BiMix}\\
A'^3_i &= \sin\theta_WA_i^\text{em} + \cos\theta_WZ_i,\label{eqn:AiMix}
\end{align}
where 
	\begin{equation}
	gA^{'3}_i=\frac{n}{r}\left(\psi_2h_2-R_2h_1-\frac{1-\cos\alpha}{\sin\theta}\right)\hat{\phi}_i,
	\end{equation}
obtained by applying the unitary gauge:
	\begin{align}
	G&=-i
		\begin{pmatrix}
		\cos\frac{\alpha}{2}&\sin\frac{\alpha}{2}e^{-in\phi}\\
		\sin\frac{\alpha}{2}e^{in\phi}&-\cos\frac{\alpha}{2}
		\end{pmatrix}\nonumber\\
	&=\cos\frac{-\pi}{2}+i\hat{u}^a_r\sigma^a\sin\frac{-\pi}{2},\nonumber\\
	\hat{u}^a_r&=\sin\frac{\alpha}{2}\cos{n\phi}\,\delta^a_1+\sin\frac{\alpha}{2}\sin{n\phi}\,\delta^a_2+\cos\frac{\alpha}{2}\,\delta^a_3.\label{eqn:gauge}
	\end{align}
In this study, we interpret constituents of $T_{33}$ through physical field combinations (Eqs. \ref{eqn:BiMix} \& \ref{eqn:AiMix}), rather than fundamental field decompositions.

Finally, to further elucidate the repulsive mechanisms, particularly the role of the $Z$-boson field, we also compute the neutral charge distribution $\mathcal{N}$ of the configuration, defined as:
	\begin{align}
	\mathcal{N}&=\partial^i\mathcal{B}^\text{neutral}_i=\partial^i\left(-\frac{1}{2}\varepsilon_{ijk}\mathcal{Z}^{jk}\right)\nonumber\\
	&=\partial^i\left[-\frac{1}{2}\varepsilon_{ijk}\left(\partial^jZ^k-\partial^kZ^j\right)\right],\label{eqn:N}
	\end{align}
where $\mathcal{B}^\text{neutral}_i$, $\mathcal{Z}^{jk}$, and $Z^j$ (as in Eq. \ref{eqn:WSMix}), are the neutral field, strength tensor, and potential, respectively.

\section{Results and Discussion}
\subsection{The Anomalous Scaling of Pole Separation}
Our investigation of the Cho-Maison MAP configurations reveals a profound departure from classical expectations of electromagnetic interaction. It is manifested in the anomalous scaling of the pole separation $d_z$, which is measured as the distance between the minima of the Higgs modulus $|\Phi|$.

Figure \ref{fig:HiggsModulus} compares the 3D surface plots of $|\Phi|$ for an SU(2) MAP and a Cho-Maison MAP in cylindrical coordinate system $(z,\rho)$, where $\rho=\sqrt{x^2+y^2}$. The behavior of $|\Phi|$ along the $z$-axis is entirely determined by $\Phi_1$, since the boundary conditions (Eq. \ref{eqn:BCtheta}) require $\Phi_2(x_c,0)=\Phi_2(x_c,\pi)=0$. The surface plot of profile function $\Phi_1$ and its cross-section $\Phi_1(x_c,0)$ for the Cho-Maison MAP (Fig. \ref{fig:HiggsModulus}(b)) are shown in Fig. \ref{fig:Phi1}. The magnetic monopole is located where $\Phi_1(x_c,0)=0$, labelled as point $C$ in Fig. \ref{fig:Phi1}(b). Similarly, the antimonopole is found at $\Phi_1(x_c,\pi)=0$. Due to the symmetry of the configuration, $d_z$ can be calculated as $2\times AC$ using the non-compactified radial coordinate $x$.

\begin{figure}
	\includegraphics[width=\columnwidth]{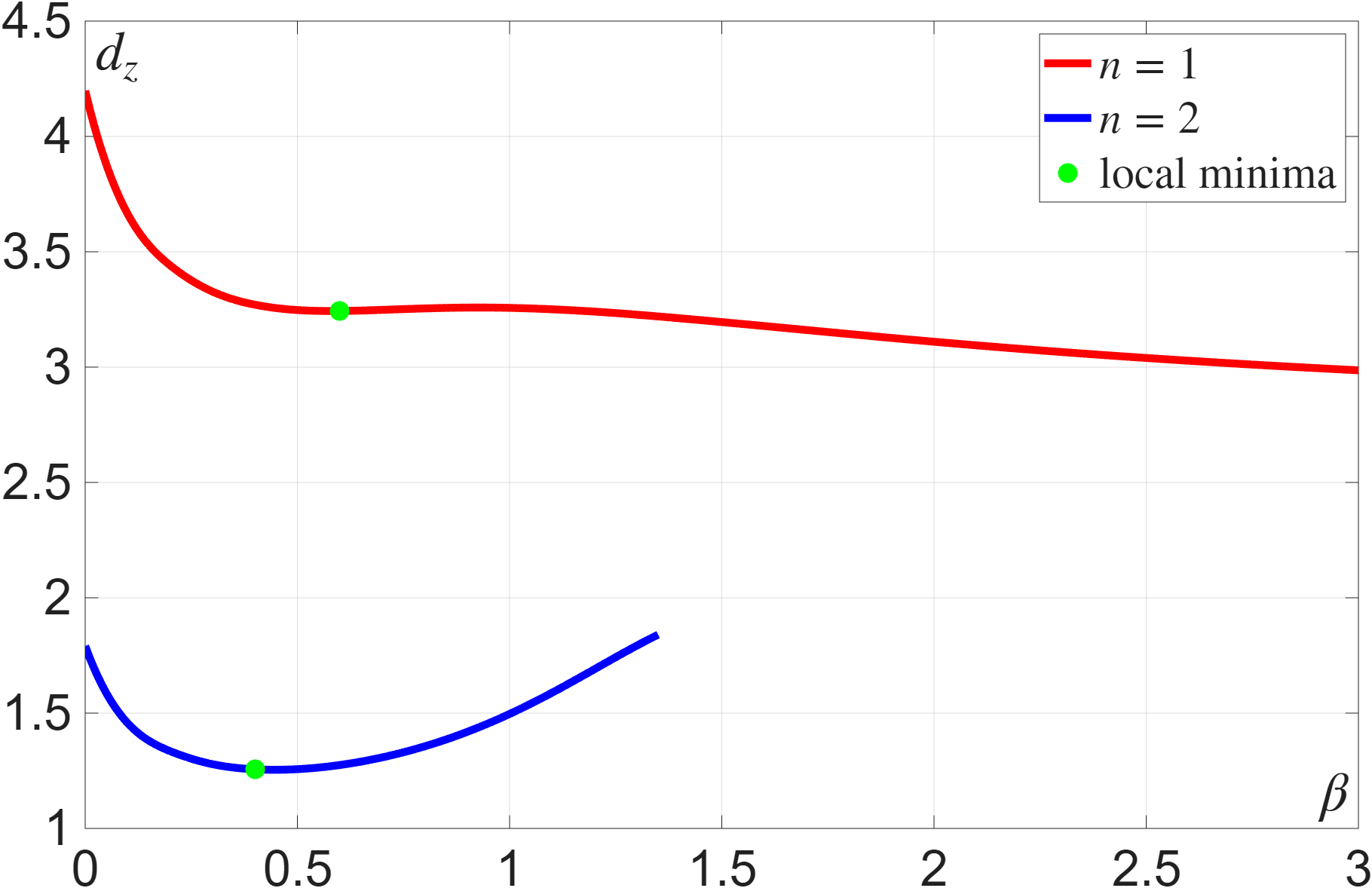}
	\caption{Pole separation $d_z$ vs. Higgs self-coupling $\beta$ for SU(2) MAP solutions with different $\phi$-winding number $n$.}
	\label{fig:dzvsbetaSU2}
\end{figure}

Intuitively, an increase in $\phi$-winding number, $n$, would strengthen the topological attraction, thereby reducing the pole separation. For the SU(2) MAPs, this trend is indeed observed, as evidenced by the variation of $d_z$ with Higgs self-coupling $\beta$ for solutions with different $n$, shown in Fig. \ref{fig:dzvsbetaSU2}.
Note that when $n=2$ (blue curve), numerical stability deteriorates when $\beta>1.35$, marked by a sharp increase in residual norm and first-order optimality; these points are consequently excluded from the plot. Furthermore, when $n\geq3$, vortex-ring formation \cite{Amin} alters the topology and $d_z$ is no longer defined.

\begin{figure}
	\includegraphics[width=\columnwidth]{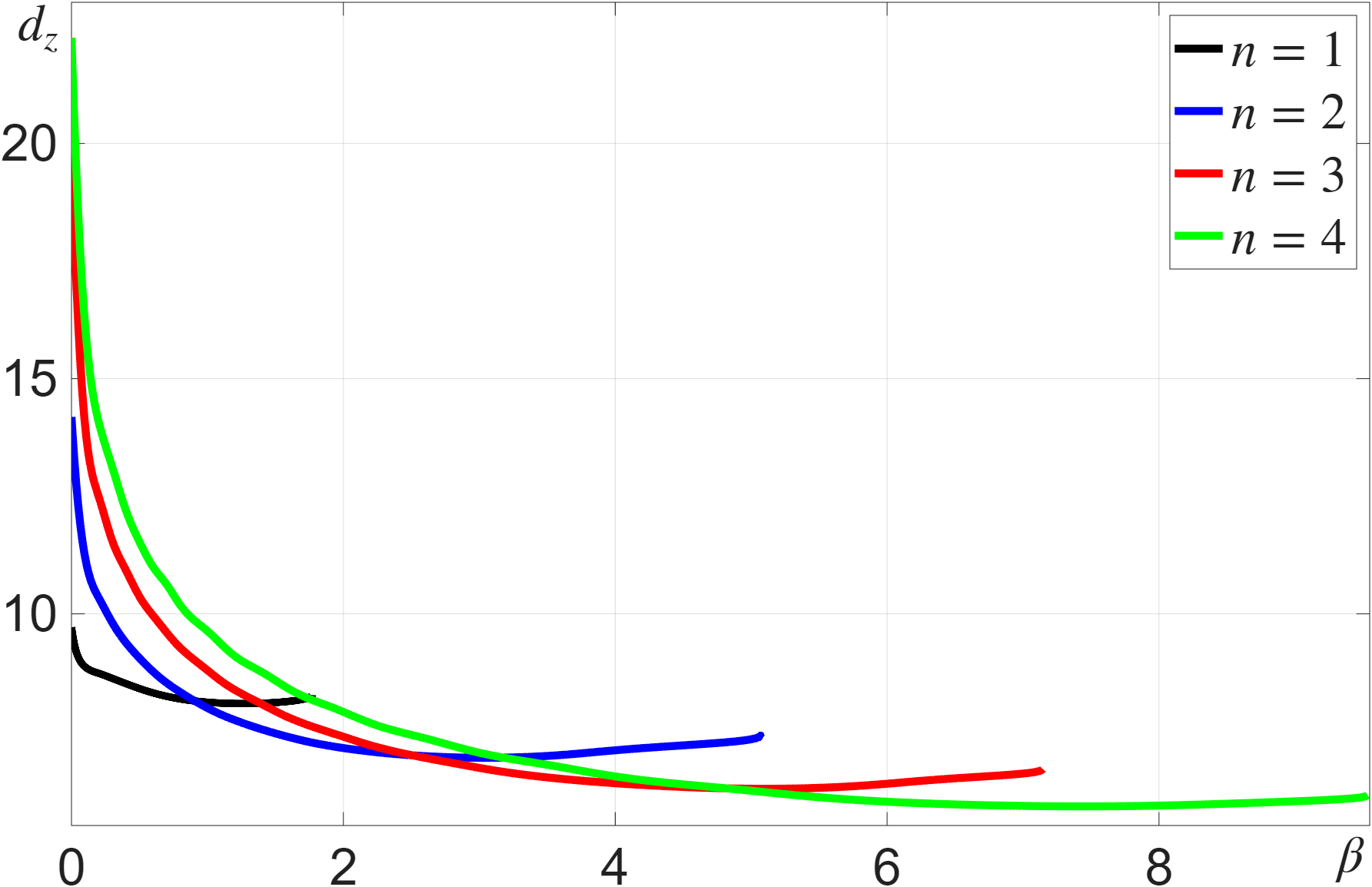}
	\caption{Pole separation $d_z$ vs. Higgs self-coupling $\beta$ for Cho-Maison MAP solutions with different $\phi$-winding number $n$.}
	\label{fig:dzvsbeta}
\end{figure}

In contrast, Fig. \ref{fig:dzvsbeta} reveals the anomalous scaling of $d_z$ versus $\beta$ for Cho-Maison MAPs with different $n$.
A striking deviation from classical expectations emerges when $\beta<1$, where $d_{z(n=1)}$ is unexpectedly the smallest. Specifically, at $\beta=0.906$, the $n=2$ curve (blue) intersects the $n=1$ curve (black) under quadrupled magnetic attraction. This implies the presence of a repulsive interaction that scales with the doubled topological charge exactly four times. As $\beta$ increases, however, the scaling systematically reverts to the expected behavior. This transition necessitates a Higgs-mediated repulsive mechanism that depends on both the topological charge and Higgs boson mass ($m_H\propto\beta$).

\subsection{The Higgs Repulsion}
While the inverse correlation between pole separation and Higgs self-coupling is well-established \cite{KKMAP,Amin}, Figs. \ref{fig:dzvsbetaSU2} \& \ref{fig:dzvsbeta} reveal a previously overlooked non-monotonicity: all curves exhibit a ``recovery phase" where $d_z$ increases with $\beta$ after reaching a minimum---most clearly seen in the minima indicated in Fig. \ref{fig:dzvsbetaSU2}. This phenomenon cannot be explained by the exponential form of a Yukawa potential. If the Higgs-mediated repulsive interaction obeys such a potential ($\sim e^{-m_Hr}/r$), the curves for $d_z$ versus $\beta$ would decrease monotonically.

	\begin{table}
	\caption{Pole separation $d_z$ for Cho-Maison MAP configurations at physical Higgs self-coupling $\beta=0.7782$ and Weinberg angle $\tan\theta_W=0.5356$ with different topological charges $n$.}
	\centering
	\begin{tabular*}{\columnwidth}{@{\extracolsep{\fill}}c c c c c c c c}
	\hline
	$n$	&1	&2	&3	&4	&5	&6	&7\\
	\hline
	$d_z$	&8.20	&8.38	&9.35	&10.30	&10.14	&9.81	&8.73\\
	\hline
	\end{tabular*}
	\label{table:1}
	\end{table}

To further probe the nature of this Higgs-mediated repulsive interaction, we push the topological complexity of this configuration to $n=7$, while fixing $\beta$, $\tan\theta_W$ at the physical values (Table \ref{table:1}). Numerical solutions cease to converge for $n\geq8$, marking the computational stability limit for this configuration. In Table \ref{table:1}, the pole separation exhibits a mirrored non-monotonicity in $n$, peaking at $n=4$—revealing the topological charge threshold where repulsive dominance reverses. This dual non-monotonicity (with respect to both $\beta$ and $n$) reveals complex Higgs-mediated repulsion dynamics and establishes it as a multi-scale phenomenon.

\begin{figure*}
	\includegraphics[width=\textwidth]{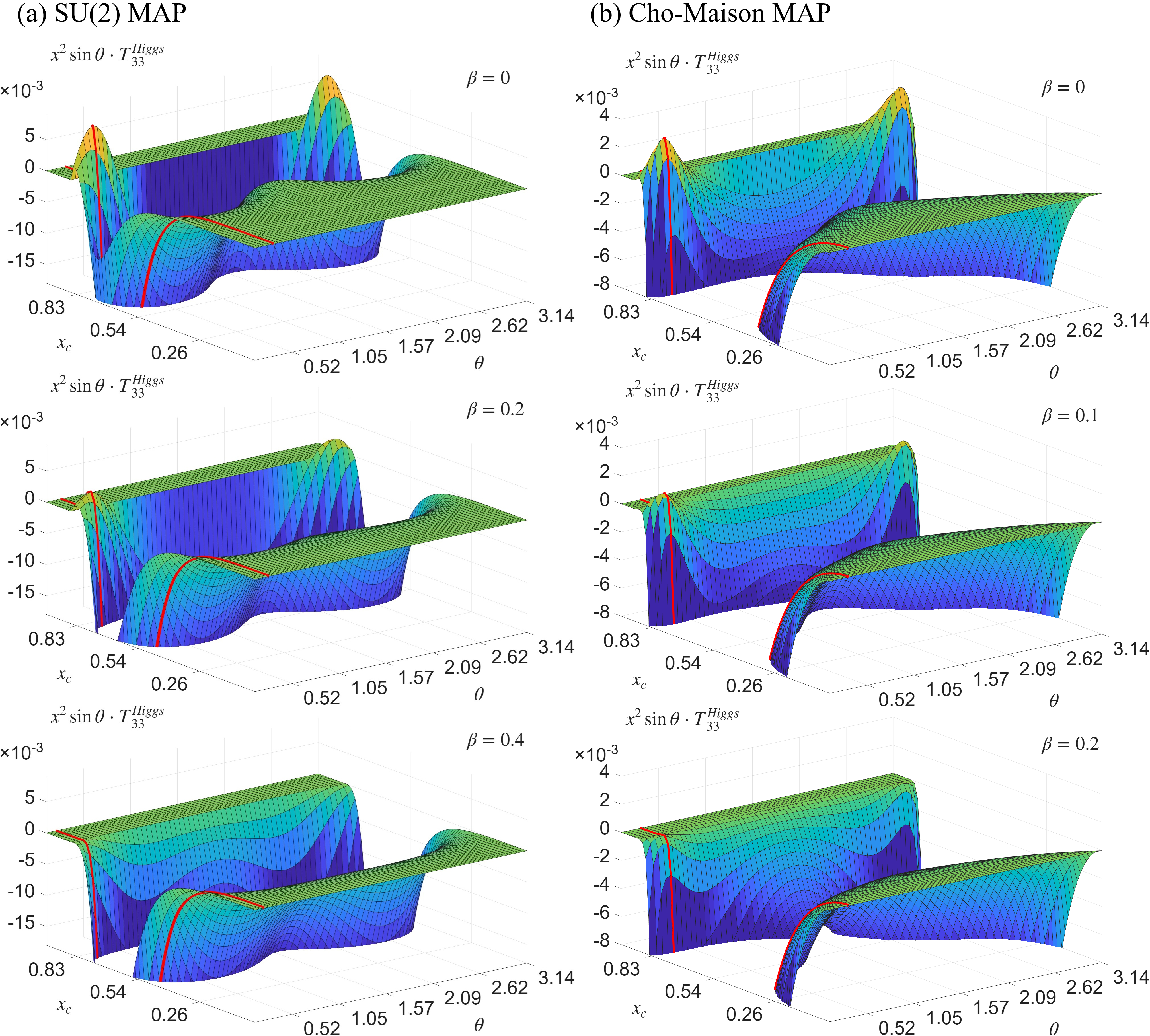}
	\caption{Evolution of the asymptotic repulsive interaction (large $x_c$) in $x^2\sin\theta\cdot T_{33}^{\text{Higgs}}$ with increasing Higgs self-coupling $\beta$ for both (a) an SU(2) MAP and (b) a Cho-Maison MAP. Note that their qualitative behaivors are identical.}
	\label{fig:t33higgsenlarged}
\end{figure*}

The anomalous scaling and the dual non-monotonic dependence together signal an intricate balance between attractive and repulsive interactions in the Cho-Maison MAP system. To unveil the underlying mechanism of this balance, we analyze the stress-energy tensor component $T_{33}$, which encodes the pressure distribution along the $z$-axis. Specifically, we investigate the Higgs and SU(2) contributions to $T_{33}$. For the U(1) contribution, however, the physics is obscured by singularities that cannot be alleviated by a weighting factor.

Figure \ref{fig:t33higgsenlarged} presents the evolution of $x^2\sin\theta\cdot T_{33}^{\text{Higgs}}$ with increasing $\beta$ for both MAP configurations, the Weinberg angle is fixed at $\tan\theta_W=0.5356$ for the Cho-Maison case. To interpret the plots, we examine the mathematical structure of $T_{33}^{\text{Higgs}}$. Derived from Eq. \ref{eqn:curlyDmu}, it comprises:
	\begin{enumerate}
		\item Direct Higgs contribution: $\left(\partial_3\bm\phi\right)^2$
		\item Gauge-Higgs couplings: $\left(A^a_i\sigma^a\bm\phi\right)^2$, ...
		\item Cross terms: $\left(A^a_i\sigma^a\bm\phi^\dagger\right)\left(B^i\bm\phi\right)$, ...
	\end{enumerate}	
Hence, $T_{33}^{\text{Higgs}}$ incorporates both pure Higgs contribution and interactions between Higgs, $Z$-boson, and electromagnetic fields (Eqs. \ref{eqn:BiMix} \& \ref{eqn:AiMix}). In particular, the Coulomb-like topological magnetic attraction between the monopoles manifests as negative contributions to the constituents of $T_{33}$, and due to the long-range nature of electromagnetic interactions, this negative contribution is expected to persist in all space and decaying asymptotically to zero as $z\rightarrow\pm\infty$.

In Fig. \ref{fig:t33higgsenlarged}, the globally negative values confirm this electromagnetic dominance in both MAPs. However, a critical sign reversal occurs asymptotically (at large $x_c$), revealing the presence of a long-range repulsive interaction that overwhelms magnetic attraction in this region. The $(x_c,\theta)$ coordinate system employed here is chosen to resolve this asymptotic behavior. Subsequent $\beta$ variations demonstrate rapid suppression of the repulsive region, vanishing completely when $\beta\geq0.2$ for the Cho-Maison case. With $\tan\theta_W$ fixed, this $\beta$ variation corresponds directly to increasing the Higgs boson mass and thus, Fig. \ref{fig:t33higgsenlarged} demonstrates a conversion from long-range to short-range repulsion mediated by the Higgs sector. To rule out the possibility that the positive values in Fig. \ref{fig:t33higgsenlarged}(b) are numerical artifacts induced by the singularities in Cho-Maison $T_{33}^{\text{Higgs}}$, we compare the plots with those of an SU(2) MAP. The qualitative behaviors of the two are identical, confirming that the positive values indeed correspond to a physical repulsive interaction.

	\begin{figure}
	\includegraphics[width=\columnwidth]{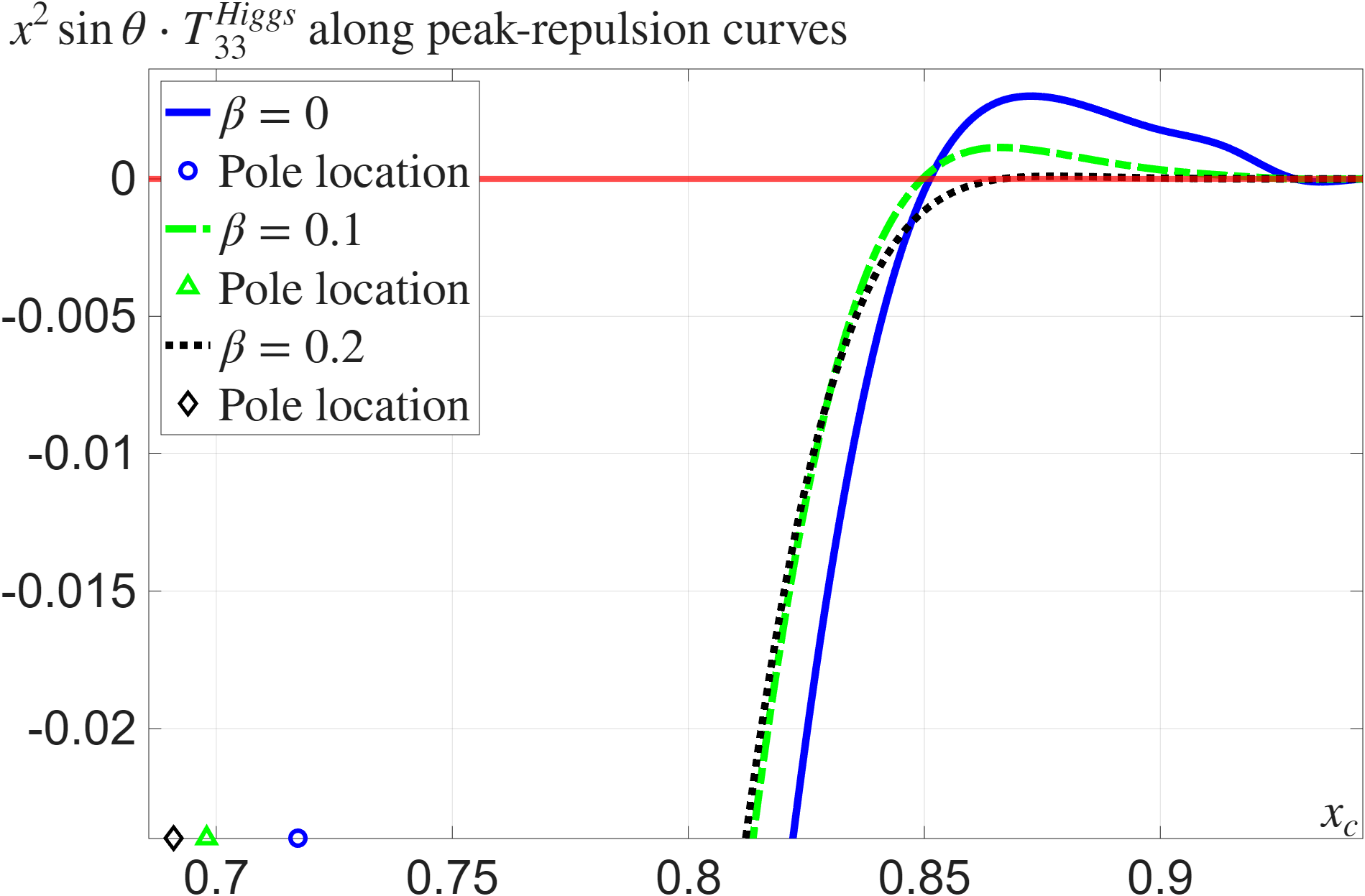}
	\caption{Cross-sections of $x^2\sin\theta\cdot T_{33}^{\text{Higgs}}$ along the peak-repulsion curves for Higgs self-coupling $\beta=$ 0, 0.1, 0.2 in the asymptotic region (large $x_c$), with pole positions marked along the bottom.}
	\label{fig:t33higgs2d}
	\end{figure}

Figure \ref{fig:t33higgs2d} displays the profiles along the peak-repulsion curves (red lines in Fig. \ref{fig:t33higgsenlarged}(b)), with the exact pole positions marked at the bottom. For $\beta=0$ (massless Higgs, solid blue curve), the pole separation is measured as $d_z=5.45\,m_W^{-1}$, representing the maximum system size. The critical sign reversal in $x^2\sin\theta\cdot T_{33}^{\text{Higgs}}$ occurs beyond $x_c=0.85$ (5.67 $m_W^{-1}$) and persists until $x_c=0.92$ (11.50 $m_W^{-1}$). The repulsive interaction thus extends to 2.11 $d_z$, confirming its long-range nature. Even at $\beta=0.1$ ($m_H=0.28\,m_W$), the Higgs repulsion remains functionally long-ranged despite having a finite mass: at $x_c=0.85$ (5.67 $m_W^{-1}$), where weak interactions are negligible, it still dominates over electromagnetic attraction. This demonstrates two key features:
	\begin{enumerate}
		\item The range transition is slower than that predicted by a Yukawa potential ($\sim e^{-m_Wr}/r$).
		\item The Higgs-mediated repulsion dominates electromagnetic attraction even at large distances and hence, its strength exceeds weak interactions by several orders of magnitude.
	\end{enumerate}
These observations suggest unconventional dynamics in the Higgs sector. Further increase in $\beta$ suppresses the repulsion completely: for $\beta=0.2$, the positive regions vanish at large $x_c$ (dotted black).

In parallel, the concurrent leftward shift of the pole positions in Fig. \ref{fig:t33higgs2d} reflects strengthened attraction. When $\beta=0$, the long-ranged Higgs repulsion completely balances electromagnetic attraction. With increasing $\beta$, the range of the repulsion shortens, leaving the electromagnetic attraction dominates at large distances. The resulting force imbalance draws the poles together until stabilized by the attenuated short-ranged Higgs repulsion.

	\begin{figure*}
	\includegraphics[width=\textwidth]{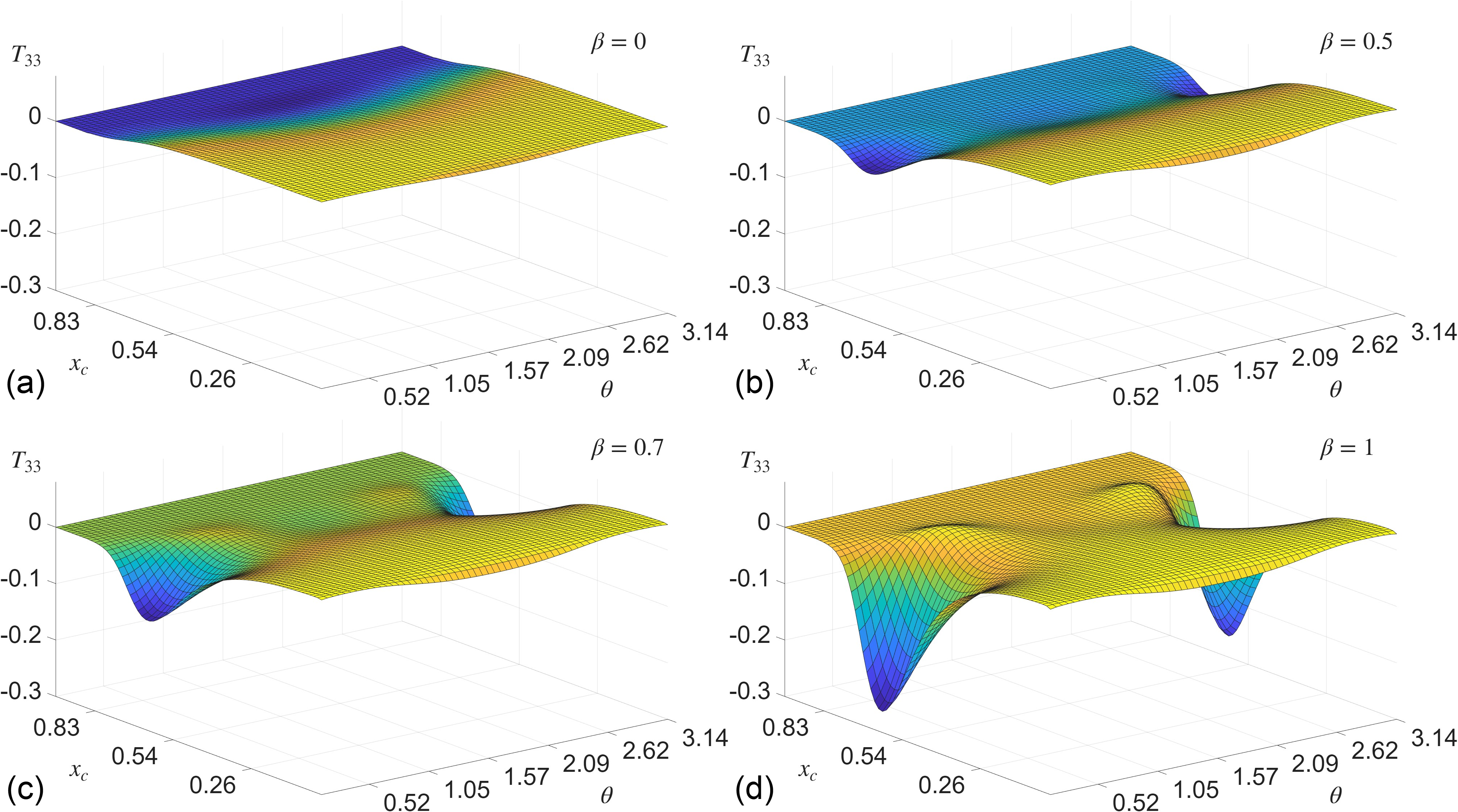}
	\caption{Evolution of the stress-energy tensor component $T_{33}$ for an SU(2) MAP with increasing Higgs self-coupling $\beta$.}
	\label{fig:t33su2evo}
	\end{figure*}

The preceding analysis of the Cho-Maison MAP configuration focuses on the weighted Higgs contribution to $T_{33}$, as the singularities prohibit the examination of the combined expression, Eq. \ref{eqn:t33}. To gain qualitative insight, we use the SU(2) Yang-Mills-Higgs theory as a benchmark, where the full expression (Eq. \ref{eqn:t33YMH}) can be examined. Figure \ref{fig:t33su2evo} shows the evolution of $T_{33}$ distribution with increasing $\beta$ for an SU(2) MAP. Remarkably, when $\beta=0$ (Fig. \ref{fig:t33su2evo}(a)), the distribution is predominantly positive in the yellow region (small $x_c$), indicating repulsive interaction dominance in the inter-pole region.

With increasing $\beta$, the Higgs-mediated repulsion becomes short-ranged and $d_z$ decreases. As a result, negative contributions to $T_{33}$ become pronounced at the monopole locations, signaling the dominance of topological magnetic attraction in these regions. Nevertheless, the overall $T_{33}$ remains complex; even when $\beta=1$ (Fig. \ref{fig:t33su2evo}(d)), positive values persist, revealing a non-trivial spatial distribution of the repulsive interaction that precludes a simple binary interpretation.

\begin{figure}
	\includegraphics[width=\columnwidth]{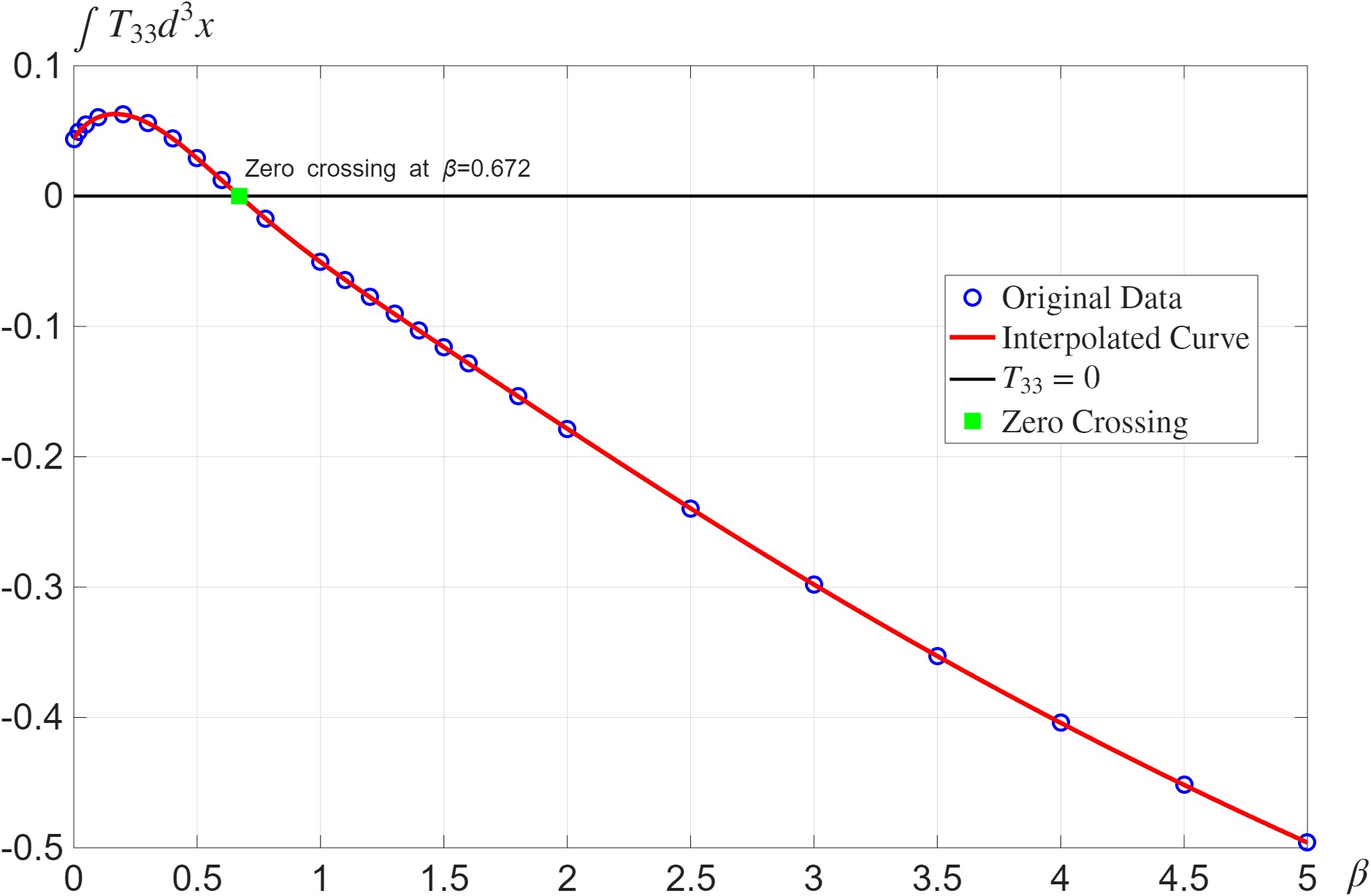}
	\caption{Plot of the integrated $T_{33}$ versus Higgs self-coupling $\beta$ for SU(2) MAP solutions with topological charge $n=1$.}
	\label{fig:t33integration}
	\end{figure}

The net stress along the $z$-axis is given by the volume integral of $T_{33}$, analogous to obtaining the total energy of the configuration from $T_{00}$. This quantity has been computed for SU(2) MAP solutions with $n=1$. Figure \ref{fig:t33integration} shows its dependence on $\beta$. A key result is the sign change in $\int T_{33}d^3x$ at $\beta=0.672$, signaling a reversal of the net force experienced from repulsive to attractive. This sign change provides definitive evidence of a competition between a Higgs-mediated repulsive interaction and the topological magnetic attraction. Since $\beta$ is the sole free parameter in the SU(2) Yang-Mills-Higgs theory and directly correlates to the Higgs boson mass, Fig. \ref{fig:t33integration} conclusively demonstrates that the Higgs field mediates a repulsive interaction in this theory. The regime $\beta<0.672$ is dominated by Higgs repulsion, pushing the monopole and antimonopole apart, whereas for $\beta>0.672$, topological magnetic attraction dominates, pulling them together toward annihilation. Together with the identical qualitative behavior of the weighted $T_{33}^{\text{Higgs}}$ (Fig. \ref{fig:t33higgsenlarged}), these findings strongly suggest that the Higgs field plays a similar repulsive role in the Weinberg-Salam model.

\subsection{The \textit{Z}-Boson Repulsion}
The force balance in the Cho-Maison MAP configuration also involves a significant contribution from the $Z$-boson field, which we analyze via $T_{33}^{\text{SU(2)}}$ in Eq. \ref{eqn:t33}. Unlike the Higgs contribution, this term is fully regular and can be visualized directly without coordinate-dependent weighting factors.

\begin{figure}
	\includegraphics[width=\columnwidth]{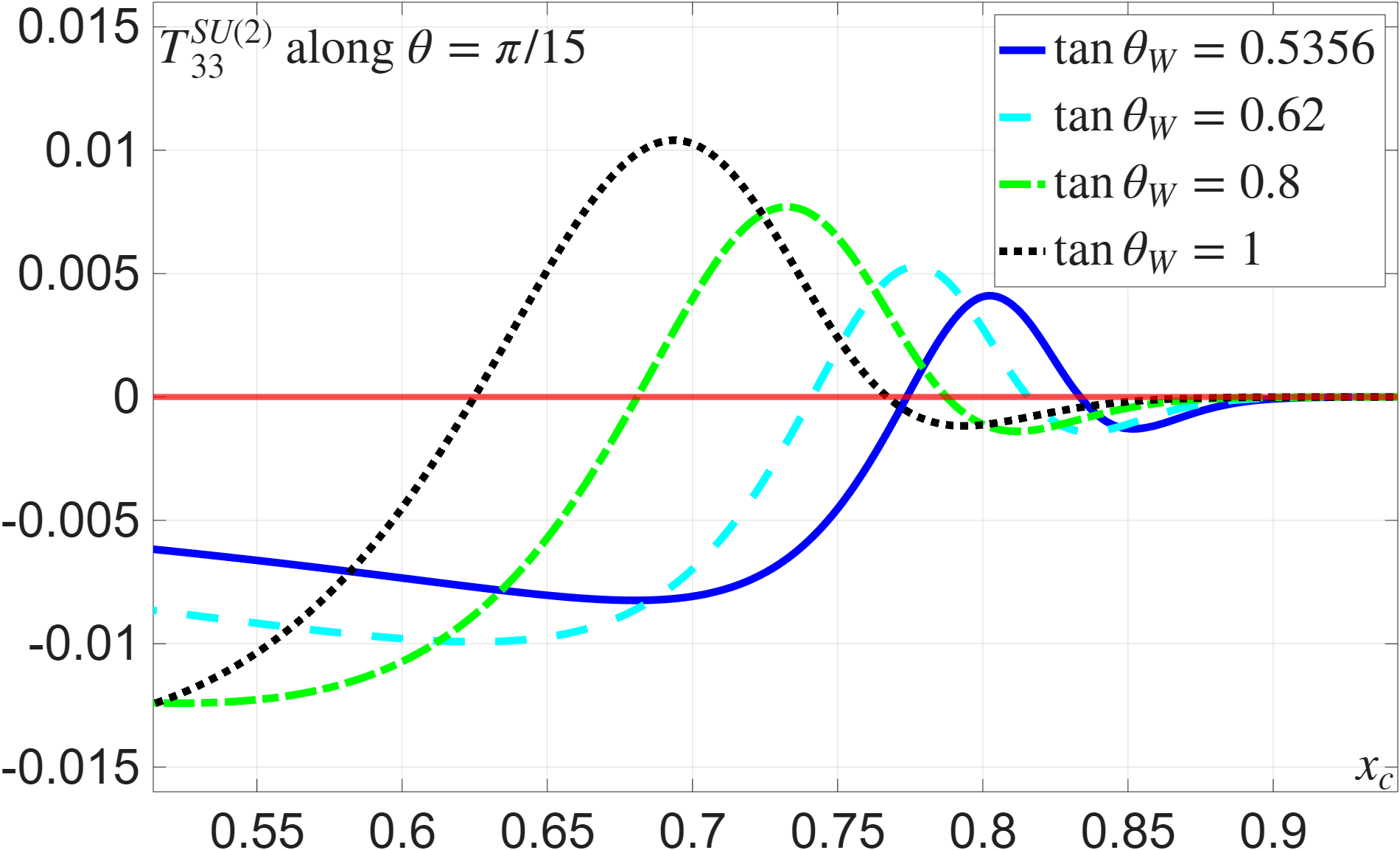}
	\caption{Cross-sections of $T_{33}^{\text{SU(2)}}$ along $\theta=\pi/15$ for solutions with physical Higgs self-coupling $\beta=0.7782$ and varying Weinberg angle $\tan\theta_W$. Repulsive interactions are localized near the pole positions.}
	\label{fig:t33su2}
	\end{figure}

Figure \ref{fig:t33su2} shows cross-sections of $T_{33}^{\text{SU(2)}}$ along $\theta=\pi/15$ (location of peak repulsion) for Cho-Maison MAP solutions with fixed $\beta$ and varying $\tan\theta_W$. The profiles reveal a globally attractive background with sharply localized repulsive cores at the pole positions. Three distinct regions can be identified in Fig. \ref{fig:t33su2} as $\tan\theta_W$ increases: (1) Enhanced electromagnetic attraction in the inter-pole region due to shortened $d_z$ (left of the positive repulsive cores); (2) Amplified repulsive core strength at the pole positions; and (3) Asymptotic stress invariance (unchanged negative $T_{33}^{\text{SU(2)}}$ at large $x_c$), indicating a persistent force balance independent of the variations in $\tan\theta_W$ and $d_z$.

The mathematical structure of Cho-Maison $T_{33}^{\text{SU(2)}}$ encodes interactions mediated by both the electromagnetic and $Z$-boson fields (Eqs. \ref{eqn:BiMix} \& \ref{eqn:AiMix}). An increase in $\tan\theta_W$ raises the $Z$-boson mass ($m_Z = m_W/\cos\theta_W$), which intensifies the weak interaction strength while shortening its effective range. This mechanism is analogous to the Higgs-mediated repulsion: as $\tan\theta_W$ increases, the range of the $Z$-boson repulsion shortens, leaving the electromagnetic attraction dominates at large distances. The resulting force imbalance draws the poles together (moving towards the left). Figure \ref{fig:t33su2} thus confirms that the $Z$-boson field provides an additional repulsive mechanism alongside the Higgs sector.

The repulsive core sizes, measured as the separation between the two zero-crossings along each curve, appear to enlarge with increasing Weinberg angle. In reality, however, the cores remain nearly identical, with a constant separation of about $1.6\,m_\text{W}^{-1}$. The apparent variation in sizes is a visual artifact of the compactified radial coordinate $x_c$. The true repulsive cores radius is therefore $R_c\approx0.8\,m_W^{-1}$---consistent with the characteristic range of weak interactions ($\sim1\ m_W^{-1}$).

	\begin{figure}
	\includegraphics[width=\columnwidth]{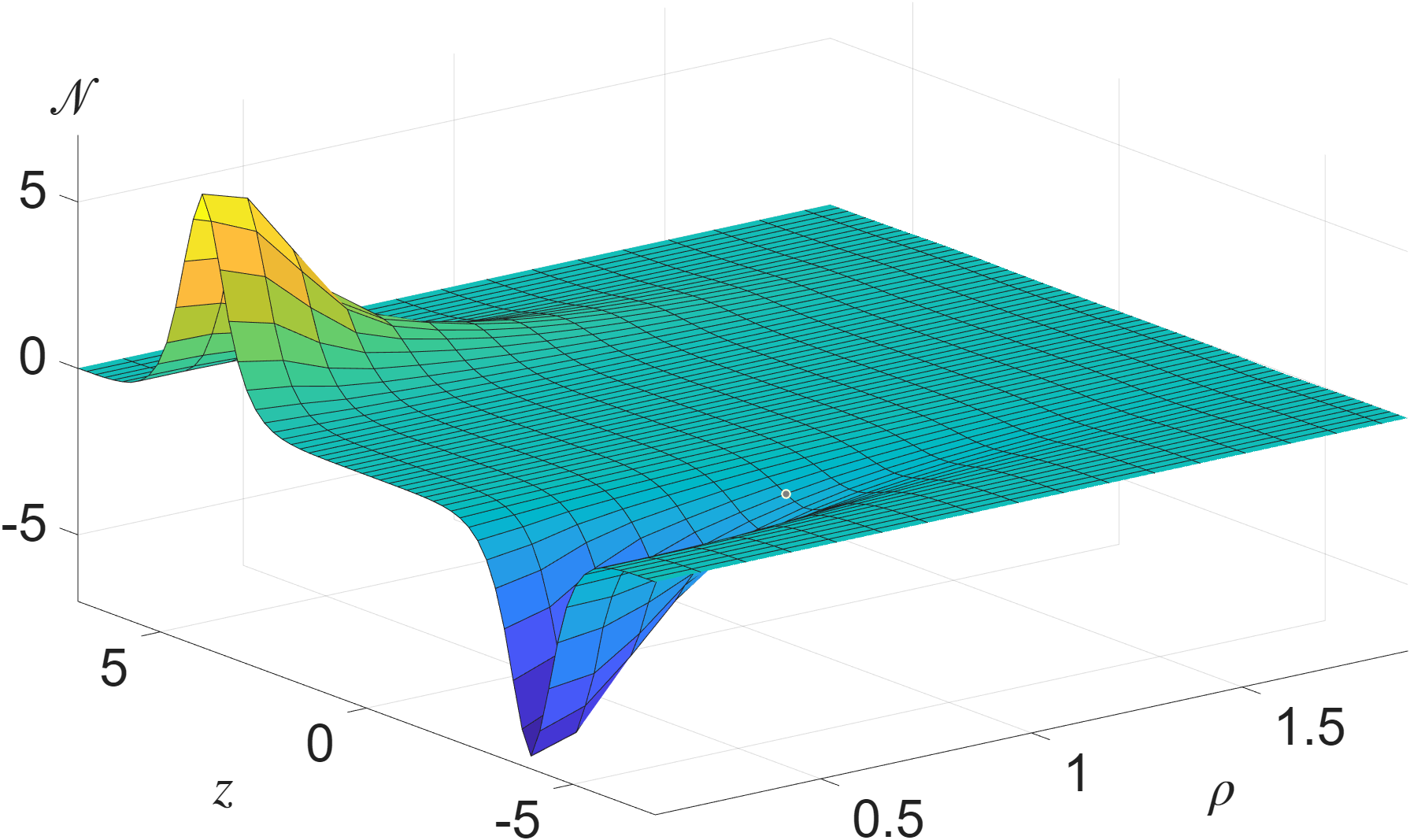}
	\caption{Cylindrical view of neutral charge density, $\mathcal{N}$, for the Cho-Maison MAP solution with $\beta = 0.7782$, $\tan\theta_W=0.5356$, and $n=1$.}
	\label{fig:Zchargedensity}
	\end{figure}

Finally, to provide a direct verification of the mediation mechanism, we compute the neutral charge density defined in Eq. \ref{eqn:N}. While an isolated Cho-Maison monopole exhibits a null neutral charge distribution \cite{ChoMaison}, Fig. \ref{fig:Zchargedensity} reveals a clear, emergent, non-zero $\mathcal{N}$ in the MAP configuration. This non-trivial charge distribution unambiguously establishes $Z$-boson exchange as the underlying mechanism responsible for generating the repulsive cores observed in Fig. \ref{fig:t33su2}.

\section{Conclusion}
This study establishes the existence of two distinct repulsive mechanisms within the Cho-Maison MAP configuration, which jointly counteract the topological magnetic attraction and are responsible for the observed finite pole separation, $d_z$:
\begin{enumerate}
    \item \textbf{Higgs-mediated repulsion} is unveiled through the dual non-monotonic dependence of $d_z$ on both the topological charge and Higgs self-coupling. The $n$-dependence confirms its topological root, while the $\beta$-dependence reveals a complexity that deviates from the exponential form of a Yukawa potential. This behavior prompted a detailed analysis of $T_{33}^{\text{Higgs}}$. The identical qualitative behaviors of this component in both the Cho-Maison and SU(2) MAPs strongly indicate that this repulsive mechanism is a robust feature of the Higgs sector, implying its presence extends to the Weinberg-Salam model. Crucially, the magnitude of this repulsion is comparable to the Coulomb-like topological magnetic attraction, highlighting its significant role in the force balance.
    \item \textbf{\textit{Z}-boson-mediated repulsion} originates purely from the weak interactions, producing localized repulsive cores of radius $R_c\approx0.8\,m_W^{-1}$ that aligns with the characteristic weak interaction scale. The emergence of a non-trivial neutral charge density distribution in the MAP configuration---absent in the spherically symmetical single Cho-Maison monopole---provides direct evidence confirming $Z$-boson as the mediator of this short-range repulsive interaction between topological defects.
\end{enumerate}

The systematic comparison between the Cho-Maison and SU(2) MAP configurations has been instrumental in this study. The analysis of the full stress-energy tensor component $T_{33}$ in the Cho-Maison case is complicated by singularities in the U(1) gauge field. Thus, the SU(2) MAP serves an essential benchmark, providing a clear reference against which the Higgs-mediated interaction in the Weinberg-Salam model can be interpreted.

The theoretical framework established here---wherein Higgs and $Z$-boson repulsions counterbalance topological magnetic attraction---extends naturally to other non-perturbative configurations. A direct application is the study of electroweak vortex-rings, which are free of the U(1) singularities encountered in the Cho-Maison MAP, thereby permitting a complete analysis of $T_{\mu\nu}$ within the Weinberg-Salam model; results in this direction will be presented in a separate publication. More broadly, the mechanism of topological repulsion mediated by a scalar field is expected to have analogues in diverse systems described by effective field theories, such as magnon spintronics and metallic hydrogen (effectively a two-component Landau-Ginzburg superconductor). Further investigation of this universal repulsive phenomenon promises deeper connections between Higgs physics, topological solitons, and effective field theories.


\end{document}